\documentclass[twocolumn,showpacs,a4paper,prb]{revtex4}

\usepackage{xspace}
\usepackage{amsmath}
\usepackage{mathptmx}

\newif\ifpdf\ifx\pdfoutput\undefined\pdffalse\else\pdftrue\fi
\ifpdf\pdfoutput=1
  \usepackage[pdftex]{graphicx}
\else
  \usepackage{graphicx}
\fi

\newcommand{\lcmo}{\ensuremath{\mathrm{La_{2/3}Ca_{1/3}MnO_{3-\delta}}}\xspace}
\newcommand{\sto}{\ensuremath{\mathrm{SrTiO_3}}\xspace}

\begin{document}

\preprint{}

\title{Low frequency $1/f$ noise in doped manganite grain-boundary junctions}

\author{J.~B.~Philipp}
\author{L.~Alff}
\author{A.~Marx}
 \email{Achim.Marx@wmi.badw.de}
 \homepage{http://www.wmi.badw-muenchen.de}
\author{R.~Gross}
\affiliation{%
Walther-Meissner-Institut, Bayerische Akademie der Wissenschaften,
Walter-Meissner-Stra{\ss}e 8, D-85748 Garching, Germany }

\date{\today}

\begin{abstract}
  
  We have performed a systematic analysis of the low frequency $1/f$-noise in
  single grain boundary junctions in the colossal magnetoresistance material
  $\mathrm{La_{2/3}Ca_{1/3}MnO_{3-\delta}}$. The grain boundary junctions were
  formed in epitaxial \lcmo films deposited on \sto bicrystal substrates and
  show a large tunneling magnetoresistance of up to 300\% at 4.2\,K as well as
  ideal, rectangular shaped resistance versus applied magnetic field curves.
  Below the Curie temperature $T_C$ the measured $1/f$ noise is dominated by
  the grain boundary. The dependence of the noise on bias current, temperature
  and applied magnetic field gives clear evidence that the large amount of low
  frequency noise is caused by localized sites with fluctuating magnetic
  moments in a heavily disordered grain boundary region.  At 4.2\,K additional
  temporally unstable Lorentzian components show up in the noise spectra that
  are most likely caused by fluctuating clusters of interacting magnetic
  moments. Noise due to fluctuating domains in the junction electrodes is
  found to play no significant role.

\end{abstract}


\maketitle

\section{Introduction}
\label{sec:introduction}

Doped manganites have attained large interest in recent years because of the
interesting interplay of charge, spin, orbital and structural degrees of
freedom in these materials \cite{Salamon01} and their potential use in
magnetoresistive devices. It was found early that the introduction of
artificial grain boundaries (GB) into epitaxial manganite thin films leads to
localized structural distortions at the GB entailing significant modifications
of the magnetotransport properties of the GB
\cite{Gubkin93a,Hwang96,Gupta96,Mathur97,Steenbeck97,Klein99a,Gross99a,Hoefener00,Ziese99a}.
In particular, a significant increase of the low field magnetoresistance was
found. Recently, in well-defined, individual GB junctions fabricated by
depositing epitaxial manganite films on \sto bicrystal substrates, a large
two-level magnetoresistance effect with a maximum tunneling magnetoresistance
(TMR) of up to 300\% at 4.2\,K has been demonstrated at low applied fields of
about 200\,Oe \cite{Hoefener00,Philipp00}. These artificial GB junctions
showed an almost ideal two-level resistance switching behavior with sharp
transitions from the low to the high resistance state when the magnetic field
was applied within the film plane parallel to the GB barrier. Thus, manganite
GB junctions represent ferromagnetic (FM) tunnel junctions with very high TMR
values and a very simple fabrication process. On the other hand, the charge
transport mechanism across the GB barrier has not yet been unambiguously
clarified. In well-defined bicrystal GB junctions defined by growing epitaxial
manganite films on \sto bicrystal substrates, the GB barrier is formed by a
straight distorted GB interface with a width of only a few nm as shown by
transmission electron microscopy \cite{Gross99a,Gross00,Wiedenhorst00}. After
annealing in oxygen atmosphere, individual GB junctions with large TMR values
have been achieved \cite{Mathur97,Steenbeck97,Gross99a,Hoefener00,Philipp00}.

Up to now, several theoretical models have been proposed
\cite{Klein99a,Gross99a,Hoefener00,Ziese99a,Evetts98a,Coey98a,Guinea98a} to
describe the magnetotransport properties of the manganite GB junctions.
However, the proposed models are controversial and a thorough understanding of
the magnetotransport properties of the GB junctions is still lacking what is
mainly related to the unknown structural and magnetic properties of the GB
barrier. Whereas Hwang {\it et al.} \cite{Hwang96} proposed a model based on
spin-polarized tunneling between ferromagnetic grains through an insulating GB
barrier, Evetts {\it et al.} \cite{Evetts98a} proposed the polarization of the
GB region by adjacent magnetically soft grains. Later on, Guinea
\cite{Guinea98a} pointed out that probably tunneling via paramagnetic impurity
states in the GB barrier plays an important role and Ziese \cite{Ziese99a}
suggested a description of the transport characteristics of GB based on
tunneling via magnetically ordered states in the barrier. Our recent
systematic study of the magnetotransport properties of well-defined individual
bicrystal GB junctions suggested a multi-step inelastic tunneling process via
magnetic impurity states within a disordered GB barrier
\cite{Klein99a,Gross99a,Hoefener00,Zhang98}. Within this model both the
nonlinear current-voltage characteristics and the strong temperature and
voltage dependence of the tunneling magnetoresistance could be naturally
explained. In our model, strain and structural distortions at the GB interface
result in a localization of charge carriers and thereby a suppression of the
ferromagnetic double exchange resulting in an insulating GB barrier with a
large density of magnetic impurity states. We also pointed out that band
bending effects may play an important role resulting in a depletion layer at
the GB interface below the Curie temperature of the ferromagnetic grain
\cite{Gross99a}.

Here, we report on a systematic analysis of the low frequency $1/f$ noise of
individual \lcmo bicrystal GB junctions to further clarify the transport
mechanism across the GB interface. The investigation of the low frequency
$1/f$ noise properties already has proven to be a valuable tool to provide
more insight into transport mechanisms across grain boundaries in the
structurally related cuprate superconductors \cite{Marx95a,Marx99,Kemen99}.
Therefore, the detailed evaluation of the $1/f$ noise of manganite GB
junctions is highly desirable both from the basic physics and the application
point of view.

For epitaxial thin films of the doped manganites there have been several
reports on a large low frequency $1/f$ noise
\cite{Alers96,Lisauskas99,Hardner97,Rajeswari96,Rajeswari98}. In particular, a
large noise peak close to the Curie temperature $T_C$ has been interpreted in
terms of a percolative nature of the transition between charge ordered
insulating and ferromagnetic metallic states \cite{Podzorov00,Anane00}.
However, Reutler et al.\ \cite{Reutler00} showed that the noise peak and the
unusually large noise level is not an intrinsic property of the doped
manganite. They found a strong coupling between local magnetic disorder and
structural disorder introduced by strain effects due to lattice mismatch
between film and substrate. In particular, the noise peak and the large $1/f$
noise level was found to be absent in high quality, strain free epitaxial
films. Palanisami et al.\ \cite{Palanisami01} suggested two different
mechanisms for the noise in manganite films: fluctuations between metallic and
insulating phases on the one hand and magnetic orientation fluctuations
(domain wall effects) on the other hand. Non-gaussian properties of the noise
together with random telegraph signals were taken as an experimental hint to
phase segregation in the colossal magnetoresistance (CMR) materials
\cite{Raquet00,Merithew00}. On the other hand, random telegraph signals
observed close to $T_C$ were taken as evidence for a domain-wall-motion
picture of the kinetics of the responsible two-level system (TLS)
\cite{Hess01}.

In contrast to epitaxial thin films there are almost no experimental data
on the noise properties of grain boundaries in the doped manganites.
Recently, Mathieu et al.\ investigated the zero-field low frequency noise
in GB junctions in $\mathrm{La_{2/3}Ca_{1/3}MnO_{3-\delta}}$
\cite{Mathieu01} below the ferromagnetic transition temperature as well as
the magnetic field dependence of the noise. They concluded that the
low-field noise was due to the multi-domain structure neighboring the GB
i.e.\ of magnetic origin. Additional Lorentzian contributions were
attributed to thermally activated domain wall motion in the domain
configuration close to the GB.

In this article we present a systematic study of the low frequency $1/f$
noise in individual grain boundary junctions formed in
$\mathrm{La_{0.67}Ca_{0.33}MnO_3}$ films. In particular, we discuss the
dependence of the measured noise on bias current, temperature and applied
magnetic field. Our results show that below $T_C$ the noise is dominated by
the GB and not by the adjacent grains. The analysis of the noise
characteristics shows that the GB noise is due to localized states with
fluctuating magnetic moments in a strongly disordered GB barrier. At the
lowest temperatures ($\simeq 4.2$\,K) additional Lorentzian contributions
show up in the noise spectra. These Lorentzians are most likely due to an
ensemble of interacting magnetic impurity states giving rise to a
simultaneous switching of their magnetic moments.

\section{Sample preparation and experimental techniques}
\label{sec:sample-prep-exper}

To achieve well-defined individual manganite grain boundary junctions (GBJs)
first about 80\,nm thick \lcmo films were grown by pulsed laser deposition on
symmetrical [001] tilt \sto bicrystal substrates with a misorientation angle
of 24$^{\circ}$. For details of the preparation process cf.\ 
Ref.~\cite{Gross00}. The \lcmo films typically had a Curie temperature
$T_C=210$\,K. After film deposition the films were annealed \emph{ex-situ} at
950$^{\circ}$C in pure oxygen atmosphere. Then, typically 30\,$\mu$m wide
microbrigdes straddling the grain boundary as well as the current and voltage
leads are patterned into the biepitaxial \lcmo films using optical lithography
and Ar ion beam etching. For comparison, microbridges of the same spatial
dimension that are not positioned across the grain boundary were patterned
into the epitaxial film. A sketch of the sample geometry is shown in the inset
of Fig.~\ref{figure1}a. The GBJs fabricated in this way were characterized by
measuring the current-voltage characteristics (IVCs) as a function of
temperature and applied magnetic field. After the annealing process the GBJs
show an almost perfect two-level resistance switching behavior with sharp
transitions between the low and high resistance level as already has been
reported recently \cite{Philipp00}.

The noise properties of the GBJs were measured by biasing the junctions at
a constant current $I_b$ and measuring the low frequency voltage
fluctuations superimposed on the resulting junction voltage. The voltage
fluctuations were amplified by low-noise amplifiers and subsequently
processed by a digital spectrum analyzer. In this way noise spectra have
been taken in the frequency range from 1\,Hz to 100\,kHz. The measurements
were performed as a function of temperature (4.2 to 300\,K) and applied
magnetic field (up to 12\,T). The magnetic field always was applied within
in the film plane parallel to the GB barrier. Great care has been taken
of the electromagnetic shielding of the sample during the noise
measurements.

In the following we will quantify the measured voltage noise power by the
frequency independent normalized voltage noise power
\begin{equation}
  \label{eq:0a}
  \Gamma = \frac{S_V}{V^2}\times f^{\alpha} \;\; .
\end{equation}
Here, $S_V$ is the spectral density of the voltage fluctuations and the
exponent $\alpha$ usually is close to unity. Below, we will usually plot the
octave integral
\begin{equation}
  \label{eq:0}
  P_{\rm octave} = \int\limits_{f_1}^{2f_1} \; \frac{S_V}{V^2} df \;\; .
\end{equation}
For $S_V\propto 1/f$ we have $P_{\rm octave}=\Gamma\ln 2$.

\section{Experimental results and discussion}
\label{sec:exper-results-disc}

\subsection{Transport and noise data}
 \label{subsec:transport-noise}

\begin{figure}
\centering{%
  \includegraphics [width=1.00\columnwidth]{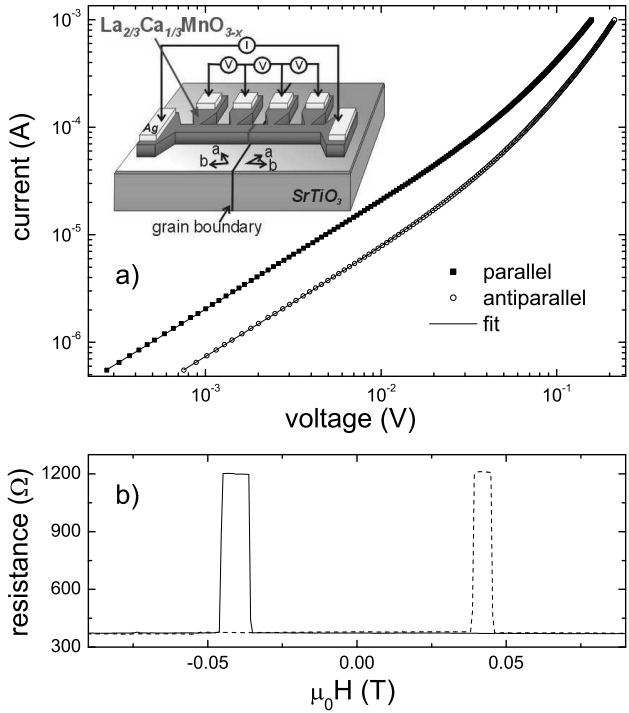}}
\caption{(a) Current-voltage characteristics of a 24$^\circ$ [001] tilt GBJ
  in a 80\,nm thick \lcmo film for parallel and antiparallel magnetization
  direction in the electrodes at $T=40$\,K. The voltage drop across the
  adjacent film parts has been subtracted. The solid lines are fits to the
  Glazman-Matveev model. The inset shows a sketch of the sample configuration.
  (b) Resistance versus applied magnetic field curve at $T=4.2$\,K showing the
  almost ideal switching behavior of the junction resistance. The field was
  applied within the field plane parallel to the grain boundary. }
 \label{figure1}
\end{figure}

We first discuss the electrical transport properties of the GBJs. Typical
current-voltage characteristics (IVCs) of a \lcmo GBJ are shown in
Fig.~\ref{figure1}a. For the parallel magnetization direction in the
electrodes the highly nonlinear IVCs can be accurately described within the
Glazman-Matveev (GM) model \cite{Glazman88} for all temperatures below
$T_C$. Within the GM model the transport of charge carriers across a
barrier containing a significant number of defect states is mediated both
by elastic tunneling (direct or resonant tunneling via a single impurity
state) and by inelastic tunneling processes via two and more defect
states. Within this model the IVCs can be expressed by
\begin{equation}
  \label{eq:1}
  I=G_1V+G_2V^{7/3}+G_3V^{7/2}+\ldots \;\; ,
\end{equation}
where $G_1$ is the elastic contribution of direct and resonant tunneling via a
single localized state and $G_2, G_3, \ldots$ give the inelastic contribution
to the total current from tunneling involving $2,3, \ldots$ impurity states.
The solid lines in Fig.~\ref{figure1}a represent fits of Eq.(\ref{eq:1}) to
the experimental data taking into account tunneling channels up to $n=3$
localized states \cite{Hoefener00}. Channels with $n>3$ are found to give only
negligible contributions. For the antiparallel magnetization configuration the
GM model also describes the IVCs at $T\simeq 40$\,K amazingly well. We note,
however, that for other temperatures the agreement with the GM prediction for
the antiparallel configuration was not as perfect as shown in
Fig.~\ref{figure1}a.

In Fig.~\ref{figure1}b we show the resistance versus applied magnetic field
curve for the magnetic field applied within the film plane parallel to the
grain boundary. As has been discussed in detail elsewhere \cite{Philipp00},
for this field direction the grain boundary junctions show an almost ideal
rectangular shaped switching behavior between the low resistance state with
parallel and the high resistance state with antiparallel magnetization
orientation in the junction electrodes.

\begin{figure}
\centering{%
  \includegraphics [width=1.00\columnwidth]{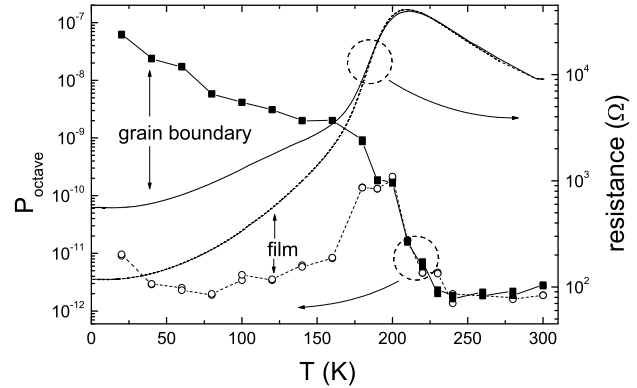}}
\caption{Temperature dependence of the normalized noise power plotted as
  octave integral from 100 to 200\,Hz for a microbridge with (full symbols,
  solid line) and without GB (open symbols, dashed line). The noise spectra
  have been taken at a sample voltage of $V=100$\,mV. For comparison the
  temperature dependence of the resistance is also shown.}
 \label{figure2}
\end{figure}

We next discuss the noise data. Figure~\ref{figure2} shows the temperature
dependence of the normalized octave noise power $P_{\rm octave}$ for two
\lcmo microbrigdes of similar geometry. Whereas one microbridge is straddling
the GB, i.e.\ contains an individual GBJ, the other is not positioned across
the GB, i.e.\ does not contain a GBJ. By comparing the noise data of these two
microbridges we can clearly identify the contribution of the GBJ to the
measured noise. The $1/f$ noise for the microbridge with the GBJ is rapidly
increasing with decreasing temperature for $T<220$\,K. In contrast, the noise
of the microbridge without GBJ is almost temperature independent except for a
peak close to $T_C$. We recently have shown that this noise peak can be
suppressed by a small applied magnetic field and is related most likely to
magnetic fluctuations at the paramagnetic to ferromagnetic transition in the
doped manganites \cite{Reutler00}. The key result of Figure~\ref{figure2} is
the fact that below $T_C$ the $1/f$ noise power of the microbridge with GBJ is
orders of magnitude larger than the noise power of the epitaxial film. That
is, for the microbridge with GBJ the measured noise can be attributed to the
GBJ alone, since the additional noise of the adjacent grains is negligible
small.

Although we do not want to discuss the details of the noise of the epitaxial
\lcmo film, we briefly compare the noise data of the epitaxial film shown in
Figure~\ref{figure2} to those reported in our previous study \cite{Reutler00}.
In Ref.~\cite{Reutler00} we have analyzed the low frequency noise in highly
strained \lcmo films grown on SrTiO$_3$ substrates. The magnitude of the noise
measured for these strained films is much larger than that measured for the
\lcmo films of our present study, although the \lcmo films were grown on the
same substrate (SrTiO$_3$) with the same lattice mismatch. These different
characteristics originate in the post-deposition annealing process applied to
the films of the present study. This annealing process results in a
significant release of the epitaxial strain and, in turn, in a reduction of
the noise amplitude. This is in agreement with our recent study, where we have
shown that the noise amplitude in strained \lcmo films is by many orders of
magnitude larger than in almost strain free films grown on NdGaO$_3$
substrates \cite{Reutler00}. The effect of a post-deposition thermal process
on the noise properties of strained manganite films has also been discussed in
Ref.~\cite{Rajeswari98}.

\begin{figure}
\centering{%
  \includegraphics [width=1.00\columnwidth]{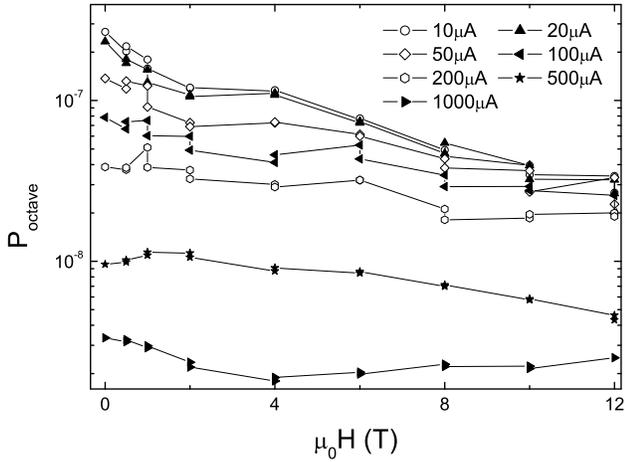}}
\caption{Magnetic field dependence of the normalized noise power for a
  \lcmo GBJ in the octave from 100 to 200\,Hz at $T=40$\,K for different
  values of the bias current $I_b$.}
 \label{figure3}
\end{figure}

Figure~\ref{figure3} shows the dependence of normalized octave noise power
$P_{\rm octave}$ on a magnetic field applied within the film plane parallel to
the GB for different values of the bias current $I_b$. Figure~\ref{figure3}
shows two experimental facts. First, the noise power decreases with increasing
bias current for all applied fields for $I_b\gtrsim 10$\,$\mu$A. Second, the
noise power decreases with increasing magnetic field for bias current values
below 500\,$\mu$A. Whereas for $I_b\lesssim 100$\,$\mu$A the noise decreases
by more than one order of magnitude by increasing the magnetic field up to
12\,T, for $I_b\gtrsim 100$\,$\mu$A the noise is only weakly dependent on the
applied magnetic field.

\begin{figure}
\centering{%
  \includegraphics [width=1.00\columnwidth]{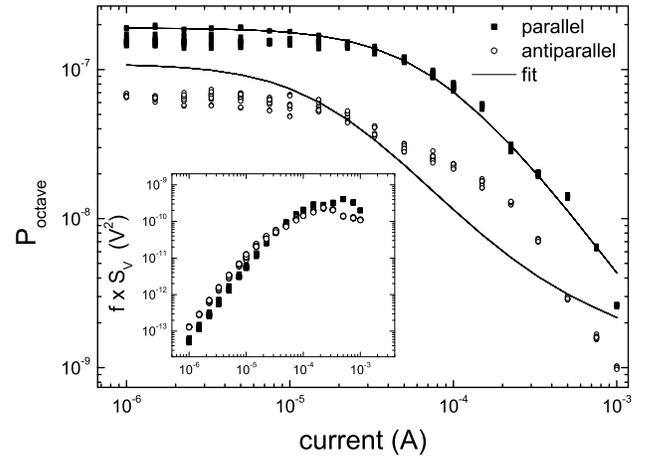}}
\caption{Normalized noise power $P_{\rm octave}$ in the octave from 100 to
  200\,Hz plotted versus the bias current $I_b$ for parallel (full symbols)
  and antiparallel magnetization direction in the GBJ electrodes (open
  symbols) at $T=40$\,K. The solid lines are fits to the data according to the
  small signal analysis (cf.\ Eq.~(\ref{eq:3})). The inset shows the noise
  power $f\times S_V$ vs the bias current $I_b$.}
 \label{figure4}
\end{figure}

Figure~\ref{figure4} shows the detailed dependence of the normalized noise
power $P_{\rm octave}$ on the bias current $I_b$ for both the parallel and
antiparallel magnetization direction in the GBJ electrodes at $T=40$\,K. It is
evident that both for the parallel and antiparallel magnetization orientation
there is only a very weak bias current dependence of the normalized noise
power for small bias currents followed by a rapid decrease of $P_{\rm octave}$
at large bias current values. As illustrated in the inset of
Fig.~\ref{figure4} the noise power $f\times S_V(I_b)$ shows a nonlinear
dependence on the bias current.

\subsection{Model considerations}
 \label{subsec:models}

\begin{figure}
\parbox{.45\columnwidth}{%
  \includegraphics [width=.5\columnwidth]{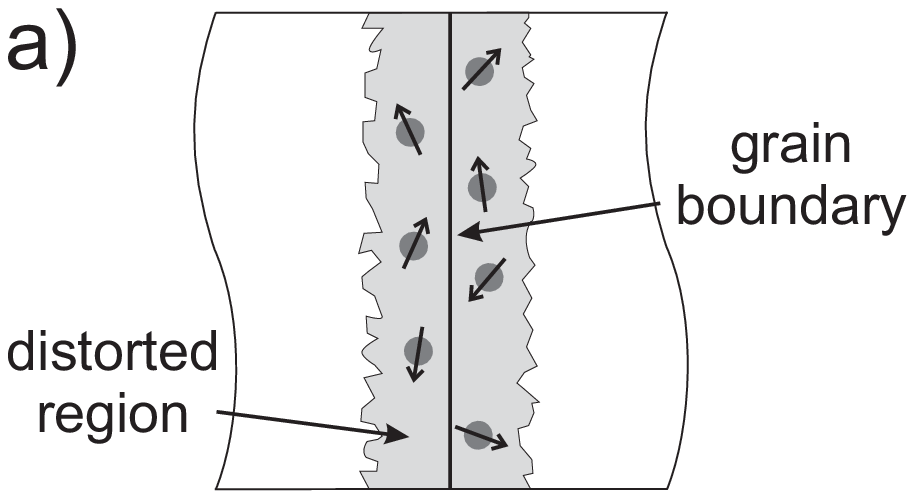}}\hfil
\parbox{.55\columnwidth}{%
  \includegraphics [width=.44\columnwidth]{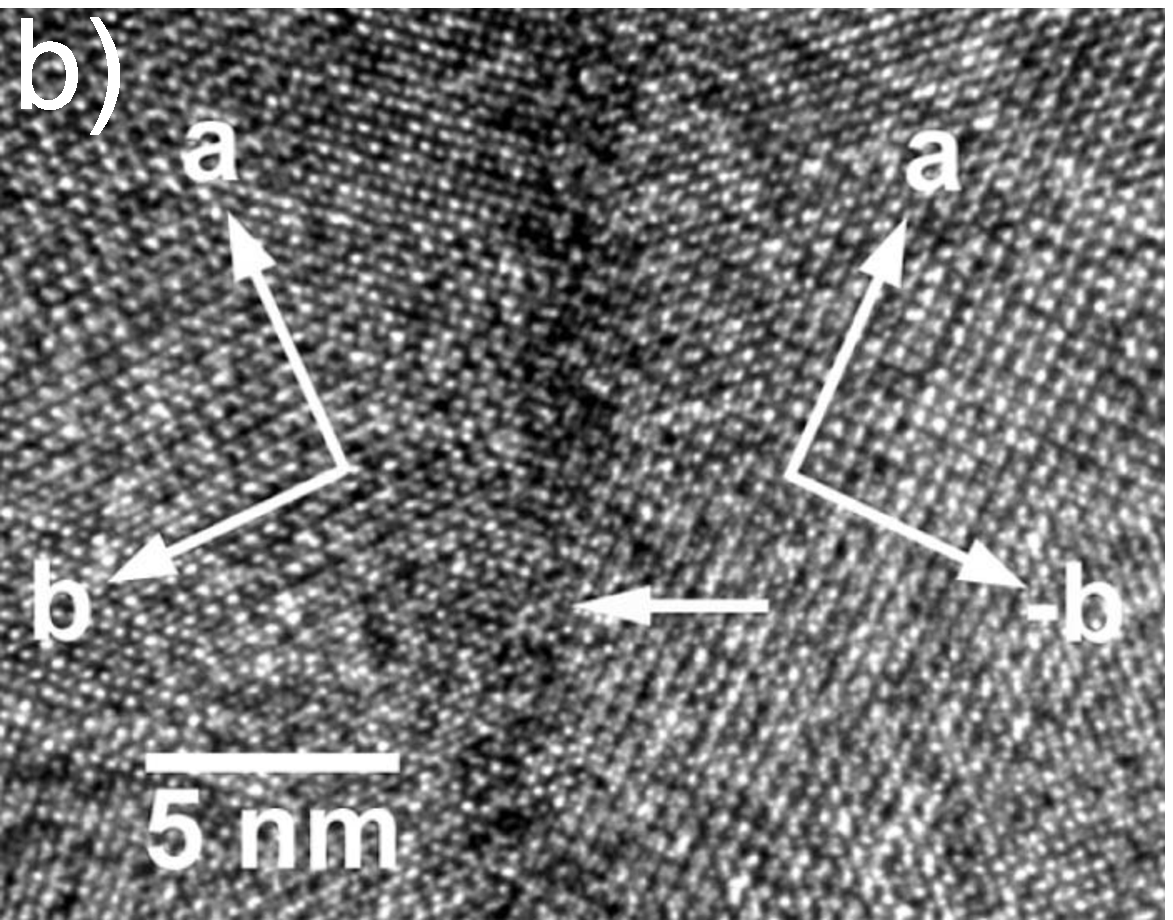}}
\caption{(a) Sketch of the junction model for grain boundary junctions in the
  doped manganites. The grain boundary interface is formed by a few nm wide
  distorted region containing a high density of defects states carrying a
  magnetic moment. The transport is by elastic tunneling as well as by
  inelastic tunneling. (b) High resolution transmission electron micrograph
  (planar view) of a grain boundary in an epitaxial \lcmo thin film grown on a
  36.8$^\circ$ symmetrical [001] tilt SrTiO$_3$ substrate. The image was
  obtained for a grain boundary that has not been annealed after the
  deposition process. The distorted grain boundary region is confined to
  within a few lattice spacings. The arrow marks a step along the straight
  grain boundary interface.}
 \label{figure5}
\end{figure}

In the following we will argue that both the dc electrical transport and
the low frequency noise properties can be consistently understood in a
junction model assuming a strongly distorted region at the GB containing a
large number of localized states/traps with fluctuating magnetic moments. A
sketch of this junction model is shown in Figure~\ref{figure5}a.

It is well known from the study of GBs in other perovskite materials (e.g.\ 
cuprate superconductors \cite{Gross94,Gross97}) that strain, structural
disorder and oxygen deficiency are important factors having a strong impact on
the electrical transport properties. Figure~\ref{figure5}b shows a high
resolution transmission electron microscopy (HR-TEM) micograph of a
symmetrical 36.8$^\circ$ [001] tilt GB in a \lcmo film deposited on a
SrTiO$_3$ bicrystal substrate \cite{Gross00}. It is obvious that the grain
boundary region is clean without any secondary phases and that the lattice
distortions are confined to within a few lattice spacings. This is very
similar to GBs in the cuprate superconductors, where the boundaries also were
found to be clean without any secondary phases and with the lattice
distortions to be confined within 1-2 lattice spacings
\cite{Chisholm91a,Kabius94a,Seo95a}.  However, the grain boundaries in doped
manganite epitaxial films are almost as straight as the GB in the underlying
SrTiO$_3$ bicrystal substrate. This is in clear contrast to GBs in the cuprate
suberconductors that are strongly facetted \cite{Kabius94a,Seo95a}. It is very
likely that this difference is related to the different growth modes of the
cuprate and manganite thin films. Whereas the cuprates show a pronounced
island growth with the islands growing across the substrate grain boundary
resulting in strong facetting, the manganites show a molecular layer-by-layer
growth mode \cite{Gross00}. In this growth mode it is expected that the grain
boundary in the film follows exactly that of the substrate. We also have
preliminary results that the microstructure of GBs in the manganites
significantly depends on the deposition technique (e.g.  laser-MBE,
sputtering), the lattice mismatch between the film and the substrate and
post-deposition annealing processes. However, more HR-TEM work is required for
a detailed clarification of this issue.

For the cuprate GBs either a description in terms of a space-wise metal
insulator transition at the grain boundary or in terms of band bending effects
\cite{Mannhart96a,Mannhart98a} lead to a description of the GB as composed of
an insulating layer at the barrier region which due to strain and structural
disorder most likely contains a high density of localized defect states.
Because of the structural affinity of the cuprate superconductors and the
doped manganites it is very likely that also for these ferromagnetic junctions
the transport properties are determined by an insulating tunnel barrier
containing a large density of localized states \cite{Gross99a}. Furthermore,
there is already strong evidence for the presence of a significant density of
localized states in the barrier from the fact that the current-voltage
characteristics of the manganite GBJs can be very well described within the GM
model (c.f.\ fig~\ref{figure1}). Further evidence comes from the strong
temperature and voltage dependence of the low field tunneling
magnetoresistance \cite{Hoefener00}. Based on this experimental evidence we
recently have proposed that the magnetotransport in manganite GBJs is
determined by multi-step inelastic tunneling via magnetic impurity states
within a disordered insulating GB barrier \cite{Klein99a,Gross99a,Hoefener00}.

Based on the model assumption of a large number of localized states within an
insulating GB barrier there are two possible mechanisms which may be
responsible for the observed low frequency voltage fluctuations. Firstly, the
localized defect states are capable of trapping and releasing charge carriers.
This charge carrier trapping and release processes lead to local variations of
the barrier height and, thus, to fluctuations of the tunneling conductance.
Furthermore, the magnetic field dependence of the noise (c.f.\ below) suggests
that the charge traps are associated with a magnetic moment with a fluctuating
orientation. Then, both the trapping and release of the charge carriers and,
hence, the charge transport between the highly spin polarized electrodes
depend on the local magnetic moment of the charge traps. Since the trapping
and releases process depends on the relative orientation of the magnetic
moment of the trap and the electrode magnetization, fluctuations of the
direction of the magnetic moments of the charge traps strongly influence the
local barrier transparency. A second mechanism giving rise to low frequency
noise is related to coupling between the localized magnetic moments
$\mathbf{S}_L$ of the localized states and the spin $\mathbf{s}$ of the
tunneling electrons. Within the simplest approximation this coupling gives
rise to an additional potential energy $U(\alpha ) = J \mathbf{S}_L \cdot
\mathbf{s} = J S_L s \cos \alpha$, where $J$ is the coupling constant and
$\alpha$ the angle between the localized moment and the electron spin. This
additional energy, which is fluctuating with fluctuating orientation of the
localized magnetic moments can be viewed as a fluctuation of the local barrier
height.

Within the model of local barrier height fluctuations the dependence of the
normalized voltage noise power $P_{\rm octave}$ on the bias current can be
described within a small signal analysis based on the GM model. Doing so, we
assume that the voltage fluctuations are caused by fluctuations of both the
elastic and inelastic current contributions due to temporal variations of the
local barrier height. Considering fluctuations of the elastic $G_1$ and
inelastic $G_2$ and $G_3$ term, Eq.~(\ref{eq:1}) gives for the small signal
voltage fluctuation
\begin{eqnarray}
  \label{eq:2}
  \delta V & \simeq & \frac{\partial V}{\partial G_1}\delta G_1
                      + \frac{\partial V}{\partial G_2}\delta G_2
                      + \frac{\partial V}{\partial G_3}\delta G_3 + \ldots \\
           & \simeq & V \; \frac{\delta G_1}{\widetilde{G}}
                      + V^{7/3} \; \frac{\delta G_2}{\widetilde{G}}
                      + V^{7/2} \; \frac{\delta G_3}{\widetilde{G}}
  \nonumber \;\; ,
\end{eqnarray}
where $\widetilde{G} = [G_1+\frac{7}{3}G_2V^{4/3}+\frac{7}{2}G_3 V^{5/2}]$
roughly corresponds to the total tunneling conductance. For independent
fluctuations $\delta G_1$, $\delta G_2$ and $\delta G_3$, the normalized
voltage noise
\begin{equation}
  \label{eq:3}
  \frac{S_V}{V^2} = \frac{S_{G_1}}{\widetilde{G}^2}
                   + V^{8/3} \; \frac{S_{G_2}}{\widetilde{G}^2}
                   + V^{5} \; \frac{S_{G_3}}{\widetilde{G}^2}
\end{equation}
is determined by the normalized fluctuations $S_{G_1}/\widetilde{G}^2=(\delta
G_1/\widetilde{G})^2$, $S_{G_2}/\widetilde{G}^2=(\delta G_2/\widetilde{G})^2$
and $S_{G_3}/\widetilde{G}^2=(\delta G_3/\widetilde{G})^2$ of the GM
coefficients.

Analyzing Eqs~(\ref{eq:2}) and (\ref{eq:3}) we can conclude the following: At
low bias current (junction voltage) the elastic tunneling current is
dominating and we can neglect the $S_{G_2}$ and $S_{G_3}$ terms and
furthermore can use the approximation $\widetilde{G}\approx G_1$. Hence, for
low bias current we expect $S_V/V^2 \approx S_{G_1}/G_1^2$, that is, a
normalized noise power independent of the bias current (junction voltage).
With increasing bias current (junction voltage) the inelastic tunneling
contribution no longer can be neglected. This results in an increase of
$\widetilde{G}$ with increasing voltage and, hence, in an overall decrease of
the of $S_V/V^2 \propto 1/\widetilde{G}^2$, even if the noise contributions of
the inelastic channels increase with increasing voltage. As shown in
Fig.~\ref{figure4} this behavior expected from our model consideration is in
good qualitativ agreement with the measured data.

We even can go further and fit the data by Eq.~(\ref{eq:3}). The solid lines
in Fig.~\ref{figure4} are fits of Eq.~(\ref{eq:3}) to the experimental data
taking into account only fluctuations of the elastic $G_1$ and the first
inelastic $G_2$ term. That is, $S_{G_1}$ and $S_{G_2}$ have been used as fit
parameters, the term $S_{G_3}$ has been neglected to keep the number of
fitting parameters minimum. We note that the GM coefficients $G_1, G_2$ and
$G_3$ entering $\widetilde{G}$ are obtained by fitting the current-voltage
characteristics by the GM model prediction and therefore are fixed parameters
in the fit of the noise data. Figure~\ref{figure4} shows that the small signal
noise analysis based on the GM model is in good agreement with the
experimental data for parallel magnetization alignment in the junction
electrodes. The values for the normalized fluctuations in the elastic channel
$S_{G_1}/G_1^2 \simeq 10^{-7}$ obtained from the numerical fits can be
compared to the noise data of GBJs in cuprate superconductors, since $(\delta
G/G)^2=(\delta R/R)^2$, where $\delta R/R$ are the normalized junction
resistance fluctuations \cite{Marx97}. The normalized $G_1$ fluctuations
nicely follow the scaling discussed in Ref.~\cite{Marx97} for the cuprate
superconductors. This scaling behavior has been discussed in terms of a
constant density of trapping centers in the cuprate GBJs. Therefore the noise
data of the manganite GBJs give further evidence of the close similarity to
cuprate GBJs and suggest a similiar density of the noise centers in both
junction types. Furthermore, $S_{G_1}$ and $S_{G_2}$ are found to depend only
weakly on temperature in the investigated temperatur range below 80\,K..

For the antiparallel magnetization orientation in the junction electrodes the
modeling of the noise data by a small signal noise analysis based on the GM
model is less convincing. However, Fig.~\ref{figure4} shows that a similar
overall dependence of the normalized noise power on the bias current is
observed for the antiparallel magnetization orientation. This is expected if
we suppose that we can use the simple Julli\`{e}re model \cite{Julliere75a} to
describe the ferromagnetic tunnel junction. Within this model the tunneling
current is given by the tunneling matrix element and the density of states
$N_{\uparrow\downarrow}(E_F)$ of the two spin directions at the Fermi level in
the junction electrodes. Going from the parallel to the antiparallel
configuration the tunneling matrix element stays the same. However, for a
material with finite spin polarization the density of states
$N_{\uparrow},_{\downarrow}(E_F)$ for the spin-up and spin-down electrons is
changed. Since in the elastic tunneling process the spin direction is
conserved, a reduction of $G_1$ and $S_{G_1}$ is expected going from the
parallel to the antiparallel magnetization orientation. It has been shown
recently \cite{Hoefener00} that in the inelastic tunneling processes the spin
direction does not seem to be conserved in manganite GBJs.  Therefore, no
reduction of $G_2$ and $G_3$ as well as $S_{G_2}$ is expected going from the
parallel to the antiparallel magnetization orientation.  Then, according to
Eq.~(\ref{eq:3}) we expect a slightly reduced value of $S_V/V^2$ and a similar
functional dependence on the bias current (junction voltage). This is in
qualitative agreement with the experimental data. We note, however, that the
Julli\`{e}re model certainly is too simple to describe the junction behavior
in full detail. In particular, the assumption of a voltage independent density
of states for the two spin directions may be an insufficient approximation.
Recently, it has been shown that band structure effects give rise to
voltage-dependent currents that conserve spin \cite{Cabrera02}. Summarizing
our discussion we can conclude that our simple model based on a insulating
tunneling barrier containing a high density of localized defect states already
describes (at least for the parallel magnetization orientation) the measured
noise data in a sufficient way. In order to get an even better agreement more
sophisticated models have to be taken into consideration.

We now discuss the dependence of $\Gamma$ or, equivalently, $P_{\rm octave}$
on temperature. As shown by Figure~\ref{figure2}, we observe an increase of
$P_{\rm octave}$ with decreasing temperature. This is expected within our
model due to the increase of the spin polarization in the junction electrodes
with decreasing temperature \cite{Hoefener00}. In this case the fluctuating
orientation of the localized magnetic moments within the GB barrier result in
increasing fluctuations of the local barrier height. Whereas for a random
orientation of the electron spin (zero spin polarization) a change of the
direction of the local magnetic moments does not change anything and hence
does not influence the tunneling probability, for a full orientation of the
electron spins (100\% spin polarization) each orientation of the localized
magnetic moment corresponds to a different potential energy $U_L = J
\mathbf{S}_L \cdot \mathbf{s}$ and hence a different local barrier height.
That is with increasing spin polarization the fluctuations of the orientation
of the localized moments results in increasing fluctuations of the local
barrier height. We note that judging from the evaluation of the IVCs within
the GM model the average barrier transparency is almost independent of
temperature.

We next discuss the dependence of $P_{\rm octave}$ on the applied magnetic
field. Applying a magnetic field was found to continuously decrease the
junction noise up to 12\,T for bias currents $\leq 500$\,$mu$A as shown in
Fig.~\ref{figure3}. This can be explained within the proposed model in a
straightforward way. The applied magnetic field tends to align the localized
magnetic moments of the charge traps in the barrier region and thus reduces
the fluctuations of the potential energy $U_L = J \mathbf{S}_L \cdot
\mathbf{s}$ and, hence, the fluctuations of the local barrier height. We note,
however, that in order to explain the magnetic field dependence of the noise
power up to the largest applied field of 12\,T (see Fig.~\ref{figure3}) the
fluctuating magnetic moments associated with the localized states cannot be
considered as free moments but rather as (weakly) interacting moments forming
a spin glass like state. It is well known that the physics of the doped
manganites is determined by a competition of ferromagnetic double exchange and
antiferromagnetic superexchange between neighboring Mn ions, which sensitively
depends on doping as well as structural disorder and bond angles. Of course,
for bulk \lcmo the ferromagnetic double exchange is dominating. However, for
the structurally distorted GB region there is certainly a strongly suppressed
double exchange resulting in locally ferromagnetic and antiferromagnetic
exchange \cite{Ziese98}. Because of this distorted nature of the GB it is
plausible to assume that there is an arrangement of interacting magnetic
moments strongly resembling a spin glass in the GB barrier.

We finally note that a spin polarized bias current of about 100\,$\mu$A
corresponding to a current density of about $10^3$\,A/cm$^2$ flowing across
the GB may result in a nonvanishing orientation of the localized magnetic
moments. That is, in this scenario the spin polarized current is expected to
have the same effect as an applied magnetic field, namely to reduce the low
frequency noise. A reduction of the noise with increasing bias current has
indeed been observed (see Fig.~\ref{figure4}) but attributed above within the
GM model to an increase of the inelastic tunneling current with increasing
bias current (junction voltage). Since the functional form of the bias current
dependence of the noise fits well to the GM model based explanation, we
conclude that the orientation effect of the spin polarized current, although
present, is small.

\begin{figure}
\centering{%
  \includegraphics [width=1.00\columnwidth]{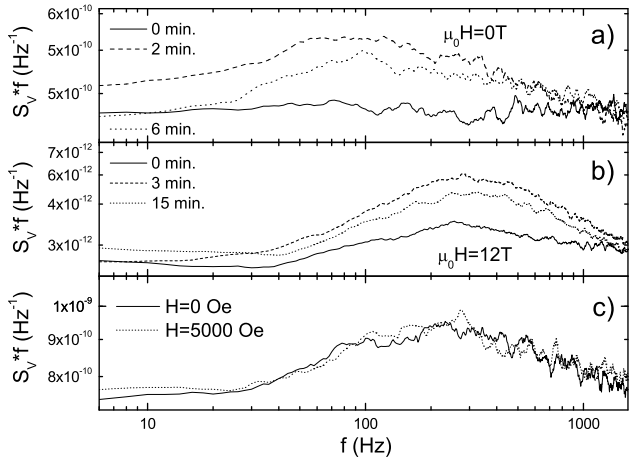}}
\caption{Voltage noise power times frequency plotted versus frequency at
  4.2\,K. Additional Lorentzian noise components are present with
  characteristic properties: some Lorentzians evolve in time at zero magnetic
  field (a) as well as in an applied magnetic field up to 12\,T (b). On the
  other hand, some Lorentzians (c) do neither change in time nor are affected
  by an applied magnetic field.}
 \label{figure6}
\end{figure}

At low temperature ($T=4.2$\,K) we observed additional Lorentzian
contributions to the low frequency $1/f$ noise as illustrated in
Fig.~\ref{figure6}. In contrast to the experiments in
Refs.~\cite{Mathieu01,Ingvarsson99,Ingvarsson00} these Lorentzians displayed
various characteristic properties that are in obvious contradiction to the
assumption of domain wall motion. First, we observe an evolution of the
Lorentzians in time both at zero magnetic field (Fig.~\ref{figure6}a) and at
an applied magnetic field of $\mu_0H=12$\,T (Fig.~\ref{figure6}b). Second, as
shown in Fig.~\ref{figure6}c) some Lorentzian components were found to be
completely unaffected by applying a magnetic field as large as several hundred
mT.  Because of this ambiguous dependence on both time and magnetic field we
suppose that the Lorentzians are due to an ensemble of interacting localized
magnetic moments. The interaction between the moments leads to simultaneous
switching of their direction between a discrete number of orientations. Random
switching of the magnetization of such an ensemble between \emph{two} distinct
directions thus defines a two level system giving rise to random telegraph
noise with a Lorentzian power spectrum. Furthermore, the independence of the
Lorentzian contribution on the magnetic field shown in Fig.~\ref{figure6} also
provides clear evidence against domain fluctuations in the junction electrodes
as the origin of GBJ noise.

\begin{figure}
\centering{%
  \includegraphics [width=1.00\columnwidth]{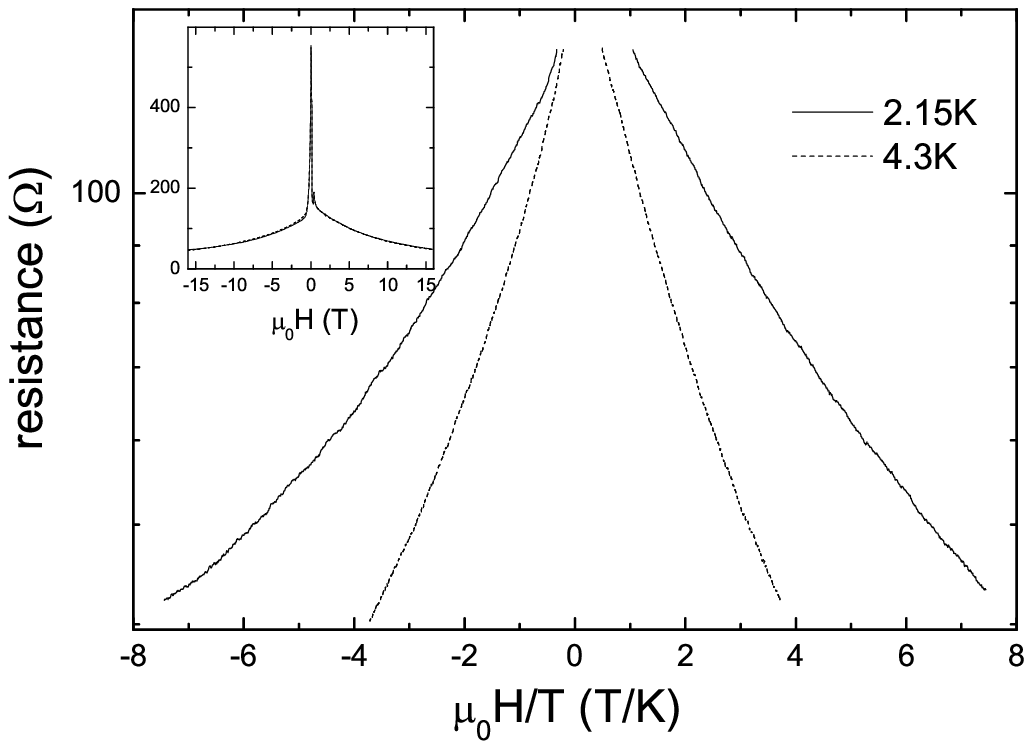}}
\caption{Resistance of a \lcmo GBJ plotted versus $\mu_0H/T$ at 2.15 and
  4.3\,K.}
 \label{figure7}
\end{figure}

To further clarify the magnetic properties of the barrier region we have
investigated the magnetic field dependence of the junction resistance down to
2\,K and up to 16\,T. Recently, in Refs.~\cite{Klein99a,Gross99a} the
distorted barrier has been modeled as a paramagnetic (PM) region even below
the Curie temperature $T_C$ of the doped manganite, since the ferromagnetic
double exchange is suppressed in the distorted GB layer.  Since according to
theoretical predictions \cite{Furukawa97} the paramagnetic insulator to
ferromagnetic metal transition in the junction electrodes is expected to
result in a large shift of the chemical potential, it has been argued that
considerable band bending effects occur at the GB interface below $T_C$. These
band bending effects lead to a depletion of the paramagnetic GB layer. The
width $t$ of this depletion layer is a function of the magnetization
difference $\Delta M = M_{\rm FM} - M_{\rm PM}$ of the FM electrodes and the
PM barrier with $t \propto \Delta M$ \cite{Klein99a,Gross99a,Furukawa97}. At
low temperature and fields above the coercive field the magnetization of the
ferromagnetic electrodes can be assumed constant and we have $t \propto \Delta
M \propto M_{\rm PM}$. The magnetization of the PM layer is determined by the
Brillouin function, which is a function of $H/T$. Then, we expect $\Delta M
\propto f(H/T)$ and, hence, $t \propto f(H/T)$. Then, the junction resistance
$R \propto \exp (-t)$ is expected to follow $R \propto \exp [-f(H/T)]$. This
is clearly not observed experimentally as shown in Fig.~\ref{figure7}, where
the two $R(H)$ curves recorded at different temperatures do not coincide when
plotted versus $H/T$. Furthermore, for $H/T \ll 1$ the Brillouin function can
be approximated by $H/T$. In this case $R \propto \exp (-H/T)$ is expected.
Such behavior has indeed been reported \cite{Klein99a} an also is shown in
Fig.~\ref{figure7}). However, for a paramagnetic GB layer at very low
temperature and very high fields the $R \propto \exp (-H/T)$ behavior should
be no longer valid, since the Brillouin function no longer can be approximated
by $H/T$. In contrast, the magnetization of the paramagnetic GB barrier is
expected to saturate resulting in a saturation of the junction resistance. As
shown by Fig.~\ref{figure7} this is clearly not observed in our experiments.
Summarizing we can conclude that the absence of a $H/T$ scaling of the
measured $R(H)$ curves and of any saturation of the junction resistance at
very high magnetic fields even at 2.3\,K provides further evidence that the
barrier region rather resembles a spin glass than a paramagnetic material in
agreement with the the above conclusions drawn from the analysis of the noise
data.

We also would briefly like to compare our noise data to that already available
in literature. Recently, Mathieu et al. \cite{Mathieu01} investigated the
$1/f$ noise of GB junctions in $\mathrm{La_{0.7}Sr_{0.3}MnO_3}$ thin films. In
these experiments the low frequency noise was found to show the same
dependence on an applied magnetic field as the dc resistance. Therefore, the
authors suggested that the measured low field noise is of magnetic origin
related to domain fluctuations and domain wall motion in a multi-domain state
neighboring the GB region. They further argued that the origin of the observed
additional Lorentzian components in the low frequency noise are caused by
thermally activated domain wall motion in this domain configuration.

Our noise data do not support this picture of a fluctuating magnetic state in
the junction electrodes adjacent to the GB. The accurate description of both
the IVCs and the noise data within the GM model strongly supports a
tunneling-like mechanism for the charge transport and gives strong evidence
that the low frequency noise in the manganite GB junctions is caused by the
trapping and release of charge carriers in localized defect states within the
GB barrier. Further support for the tunnel junction model stems from the
$R(H)$ dependencies \cite{Philipp00}. Here, for our GBJs an ideal two-level
resistance switching with sharp transition from the low to the high resistance
state is observed with the magnetic field applied parallel to the GB barrier
resembling the rectangular shaped $R(H)$ characteristics observed in TMR
devices based on transition metals \cite{Jansen00}. Such $R(H)$ dependencies
would not be expected in the presence of a multi-domain state in the junction
electrodes. Furthermore, the normalized octave noise $P_{\rm octave}$ of
Fig.~\ref{figure3} shows a strong magnetic field dependence up to applied
fields of 12\,T especially at low bias currents. Again, this in contradiction
to a multi-domain state, for which domain fluctuations are expected to be
strongly suppressed at fields above about 1\,T where the domains are fully
aligned.

\section{Summary}
\label{sec:summary}

In summary, we performed a detailed analysis of the low frequency $1/f$-noise
in individual bicrystal grain boundary junctions formed in epitaxial
{$\mathrm{La_{0.67}Ca_{0.33}MnO_3}$} films as a function of temperature, bias
current, and applied magnetic field. Our noise data show that the low
frequency noise in these junctions showing nearly ideal two-level resistance
switching is due to localized sites with fluctuating magnetic moments in a
strongly distorted barrier region. This is in full agreement with the
description of the electrical transport properties of the GBJs by elastic and
inelastic tunneling via localized defects states within an insulating grain
boundary barrier. Low frequency noise due to domain fluctuations in the
junction electrodes is found to play no significant role in the investigated
samples. Additional Lorentzian contributions to the noise showing up at low
temperature most likely are caused by clusters of interacting magnetic
moments. The analysis of the electrical transport properties and the noise up
to high magnetic fields suggests that the grain boundary barrier is rather a
spin glass than a paramagnetic layer.

\section{Acknowledgment}
\label{sec:acknowledgement}

The authors want to thank C.~H\"{o}fener and J.~Klein for valuable discussions.
This work has been supported by the BMBF.

\bibliography{Philipp2002}

\begin{thebibliography}{10}
\expandafter\ifx\csname bibnamefont\endcsname\relax
  \def\bibnamefont#1{#1}\fi
\expandafter\ifx\csname bibfnamefont\endcsname\relax
  \def\bibfnamefont#1{#1}\fi
\expandafter\ifx\csname url\endcsname\relax
  \def\url#1{\texttt{#1}}\fi
\expandafter\ifx\csname urlprefix\endcsname\relax\def\urlprefix{URL }\fi
\expandafter\ifx\csname bibinfo\endcsname\relax \def\bibinfo#1#2{#2}\fi
\expandafter\ifx\csname eprint\endcsname\relax \def\eprint#1{#1}\fi

\bibitem{Salamon01}
\bibinfo{author}{\bibfnamefont{M.~B.} \bibnamefont{Salamon}} \bibnamefont{and}
  \bibinfo{author}{\bibfnamefont{M.}~\bibnamefont{Jaime}},
  \bibinfo{journal}{Rev. Mod. Phys.} \textbf{\bibinfo{volume}{73}},
  \bibinfo{pages}{583} (\bibinfo{year}{2001}).

\bibitem{Gubkin93a}
\bibinfo{author}{\bibfnamefont{F.}~\bibnamefont{Guinea}},
  \bibinfo{journal}{Phys. Solid State} \textbf{\bibinfo{volume}{35}},
  \bibinfo{pages}{728} (\bibinfo{year}{1993}).

\bibitem{Hwang96}
\bibinfo{author}{\bibfnamefont{H.~Y.} \bibnamefont{Hwang}},
  \bibinfo{author}{\bibfnamefont{S.-W.} \bibnamefont{Cheong}},
  \bibinfo{author}{\bibfnamefont{N.~P.} \bibnamefont{Ong}}, \bibnamefont{and}
  \bibinfo{author}{\bibfnamefont{B.}~\bibnamefont{Barlogg}},
  \bibinfo{journal}{Phys. Rev. Lett.} \textbf{\bibinfo{volume}{77}},
  \bibinfo{pages}{2041} (\bibinfo{year}{1996}).

\bibitem{Gupta96}
\bibinfo{author}{\bibfnamefont{A.}~\bibnamefont{Gupta}},
  \bibinfo{author}{\bibfnamefont{G.~Q.} \bibnamefont{Gong}},
  \bibinfo{author}{\bibfnamefont{G.}~\bibnamefont{Xiao}},
  \bibinfo{author}{\bibfnamefont{P.~R.} \bibnamefont{Duncombe}},
  \bibinfo{author}{\bibfnamefont{P.}~\bibnamefont{Lecoeur}},
  \bibinfo{author}{\bibfnamefont{P.}~\bibnamefont{Trouilloud}},
  \bibinfo{author}{\bibfnamefont{Y.~Y.} \bibnamefont{Wang}},
  \bibinfo{author}{\bibfnamefont{V.~P.} \bibnamefont{Dravid}},
  \bibnamefont{and} \bibinfo{author}{\bibfnamefont{J.~Z.} \bibnamefont{Sun}},
  \bibinfo{journal}{Phys. Rev. B} \textbf{\bibinfo{volume}{54}},
  \bibinfo{pages}{R15629} (\bibinfo{year}{1996}).

\bibitem{Mathur97}
\bibinfo{author}{\bibfnamefont{N.~D.} \bibnamefont{Mathur}},
  \bibinfo{author}{\bibfnamefont{G.}~\bibnamefont{Burnell}},
  \bibinfo{author}{\bibfnamefont{S.~P.} \bibnamefont{Isaac}},
  \bibinfo{author}{\bibfnamefont{T.~J.} \bibnamefont{Jackson}},
  \bibinfo{author}{\bibfnamefont{B.-S.} \bibnamefont{Teo}},
  \bibinfo{author}{\bibfnamefont{J.~L.} \bibnamefont{{MacManus-Driscoll}}},
  \bibinfo{author}{\bibfnamefont{L.~F.} \bibnamefont{Cohen}},
  \bibinfo{author}{\bibfnamefont{J.~E.} \bibnamefont{Evetts}},
  \bibnamefont{and} \bibinfo{author}{\bibfnamefont{M.~G.}
  \bibnamefont{Blamire}}, \bibinfo{journal}{Nature}
  \textbf{\bibinfo{volume}{387}}, \bibinfo{pages}{266} (\bibinfo{year}{1997}).

\bibitem{Steenbeck97}
\bibinfo{author}{\bibfnamefont{K.}~\bibnamefont{Steenbeck}},
  \bibinfo{author}{\bibfnamefont{T.}~\bibnamefont{Eick}},
  \bibinfo{author}{\bibfnamefont{K.}~\bibnamefont{Kirsch}},
  \bibinfo{author}{\bibfnamefont{K.}~\bibnamefont{{O'Donnell}}},
  \bibnamefont{and}
  \bibinfo{author}{\bibfnamefont{E.}~\bibnamefont{Steinbei{\ss}}},
  \bibinfo{journal}{Appl. Phys. Lett.} \textbf{\bibinfo{volume}{71}},
  \bibinfo{pages}{968} (\bibinfo{year}{1997}).

\bibitem{Klein99a}
\bibinfo{author}{\bibfnamefont{J.}~\bibnamefont{Klein}},
  \bibinfo{author}{\bibfnamefont{C.}~\bibnamefont{H{\"o}fener}},
  \bibinfo{author}{\bibfnamefont{S.}~\bibnamefont{Uhlenbruck}},
  \bibinfo{author}{\bibfnamefont{L.}~\bibnamefont{Alff}},
  \bibinfo{author}{\bibfnamefont{B.}~\bibnamefont{B{\"u}chner}},
  \bibnamefont{and} \bibinfo{author}{\bibfnamefont{R.}~\bibnamefont{Gross}},
  \bibinfo{journal}{Europhys. Lett.} \textbf{\bibinfo{volume}{47}},
  \bibinfo{pages}{371} (\bibinfo{year}{1999}).

\bibitem{Gross99a}
\bibinfo{author}{\bibfnamefont{R.}~\bibnamefont{Gross}},
  \bibinfo{author}{\bibfnamefont{L.}~\bibnamefont{Alff}},
  \bibinfo{author}{\bibfnamefont{B.}~\bibnamefont{B{\"u}chner}},
  \bibinfo{author}{\bibfnamefont{B.~H.} \bibnamefont{Freitag}},
  \bibinfo{author}{\bibfnamefont{C.}~\bibnamefont{H{\"o}fener}},
  \bibinfo{author}{\bibfnamefont{J.}~\bibnamefont{Klein}},
  \bibinfo{author}{\bibfnamefont{Y.}~\bibnamefont{Lu}},
  \bibinfo{author}{\bibfnamefont{W.}~\bibnamefont{Mader}},
  \bibinfo{author}{\bibfnamefont{J.~B.} \bibnamefont{Philipp}},
  \bibinfo{author}{\bibfnamefont{M.~S.~R.} \bibnamefont{Rao}},
  \bibinfo{author}{\bibfnamefont{P.}~\bibnamefont{Reutler}},
  \bibinfo{author}{\bibfnamefont{S.}~\bibnamefont{Ritter}}, \emph{et~al.},
  \bibinfo{journal}{J. Magn. Magn. Mater.} \textbf{\bibinfo{volume}{211}},
  \bibinfo{pages}{150} (\bibinfo{year}{2000}).

\bibitem{Hoefener00}
\bibinfo{author}{\bibfnamefont{C.}~\bibnamefont{H{\"o}fener}},
  \bibinfo{author}{\bibfnamefont{J.~B.} \bibnamefont{Philipp}},
  \bibinfo{author}{\bibfnamefont{J.}~\bibnamefont{Klein}},
  \bibinfo{author}{\bibfnamefont{L.}~\bibnamefont{Alff}},
  \bibinfo{author}{\bibfnamefont{A.}~\bibnamefont{Marx}},
  \bibinfo{author}{\bibfnamefont{B.}~\bibnamefont{B{\"u}chener}},
  \bibnamefont{and} \bibinfo{author}{\bibfnamefont{R.}~\bibnamefont{Gross}},
  \bibinfo{journal}{Europhys. Lett.} \textbf{\bibinfo{volume}{50}},
  \bibinfo{pages}{681} (\bibinfo{year}{2000}).

\bibitem{Ziese99a}
\bibinfo{author}{\bibfnamefont{M.}~\bibnamefont{Ziese}},
  \bibinfo{journal}{Phys. Rev. B} \textbf{\bibinfo{volume}{60}},
  \bibinfo{pages}{R738} (\bibinfo{year}{1999}).

\bibitem{Philipp00}
\bibinfo{author}{\bibfnamefont{J.~B.} \bibnamefont{Philipp}},
  \bibinfo{author}{\bibfnamefont{C.}~\bibnamefont{H{\"o}fener}},
  \bibinfo{author}{\bibfnamefont{S.}~\bibnamefont{Thienhaus}},
  \bibinfo{author}{\bibfnamefont{J.}~\bibnamefont{Klein}},
  \bibinfo{author}{\bibfnamefont{L.}~\bibnamefont{Alff}}, \bibnamefont{and}
  \bibinfo{author}{\bibfnamefont{R.}~\bibnamefont{Gross}},
  \bibinfo{journal}{Phys. Rev. B} \textbf{\bibinfo{volume}{62}},
  \bibinfo{pages}{9248} (\bibinfo{year}{2000}).

\bibitem{Gross00}
\bibinfo{author}{\bibfnamefont{R.}~\bibnamefont{Gross}},
  \bibinfo{author}{\bibfnamefont{J.}~\bibnamefont{Klein}},
  \bibinfo{author}{\bibfnamefont{B.}~\bibnamefont{Wiedenhorst}},
  \bibinfo{author}{\bibfnamefont{C.}~\bibnamefont{H{\"o}fener}},
  \bibinfo{author}{\bibfnamefont{U.}~\bibnamefont{Schoop}},
  \bibinfo{author}{\bibfnamefont{J.~B.} \bibnamefont{Philipp}},
  \bibinfo{author}{\bibfnamefont{M.}~\bibnamefont{Schonecke}},
  \bibinfo{author}{\bibfnamefont{F.}~\bibnamefont{Herbstritt}},
  \bibinfo{author}{\bibfnamefont{L.}~\bibnamefont{Alff}},
  \bibinfo{author}{\bibfnamefont{Y.}~\bibnamefont{Lu}},
  \bibinfo{author}{\bibfnamefont{A.}~\bibnamefont{Marx}},
  \bibinfo{author}{\bibfnamefont{S.}~\bibnamefont{Schymon}}, \emph{et~al.}, in
  \emph{\bibinfo{booktitle}{Superconducting and Related Oxides: Physics and
  Nanoengineering IV}}, edited by
  \bibinfo{editor}{\bibfnamefont{D.}~\bibnamefont{Pavuna}} \bibnamefont{and}
  \bibinfo{editor}{\bibfnamefont{I.}~\bibnamefont{Bosovic}}
  (\bibinfo{year}{2000}), vol. \bibinfo{volume}{4058} of
  \emph{\bibinfo{series}{SPIE Conf. Proc.}}, pp. \bibinfo{pages}{278--294}.

\bibitem{Wiedenhorst00}
\bibinfo{author}{\bibfnamefont{B.}~\bibnamefont{Wiedenhorst}},
  \bibinfo{author}{\bibfnamefont{L.}~\bibnamefont{Alff}},
  \bibinfo{author}{\bibfnamefont{C.}~\bibnamefont{Recher}},
  \bibinfo{author}{\bibfnamefont{J.}~\bibnamefont{Klein}},
  \bibinfo{author}{\bibfnamefont{R.}~\bibnamefont{Gross}},
  \bibinfo{author}{\bibfnamefont{T.}~\bibnamefont{Walther}}, ,
  \bibnamefont{and} \bibinfo{author}{\bibfnamefont{W.}~\bibnamefont{Mader}},
  \bibinfo{journal}{J. Magn. and Magn. Mat.} \textbf{\bibinfo{volume}{211}},
  \bibinfo{pages}{16} (\bibinfo{year}{2000}).

\bibitem{Evetts98a}
\bibinfo{author}{\bibfnamefont{J.~E.} \bibnamefont{Evetts}},
  \bibinfo{author}{\bibfnamefont{M.~G.} \bibnamefont{Blamire}},
  \bibinfo{author}{\bibfnamefont{N.~D.} \bibnamefont{Mathur}},
  \bibinfo{author}{\bibfnamefont{S.~P.} \bibnamefont{Isaac}},
  \bibinfo{author}{\bibfnamefont{B.-S.} \bibnamefont{Teo}},
  \bibinfo{author}{\bibfnamefont{L.~F.} \bibnamefont{Cohem}}, \bibnamefont{and}
  \bibinfo{author}{\bibfnamefont{J.~L.} \bibnamefont{MacManus-Driscoll}},
  \bibinfo{journal}{Trans. R. Soc. Lond. A} \textbf{\bibinfo{volume}{356}},
  \bibinfo{pages}{1593} (\bibinfo{year}{1998}).

\bibitem{Coey98a}
\bibinfo{author}{\bibfnamefont{J.~M.~D.} \bibnamefont{Coey}},
  \bibinfo{author}{\bibfnamefont{A.~E.} \bibnamefont{Berkowitz}},
  \bibinfo{author}{\bibfnamefont{L.}~\bibnamefont{Balcells}},
  \bibinfo{author}{\bibfnamefont{F.~F.} \bibnamefont{Putris}},
  \bibnamefont{and} \bibinfo{author}{\bibfnamefont{A.}~\bibnamefont{Barry}},
  \bibinfo{journal}{Phys. Rev. Lett.} \textbf{\bibinfo{volume}{80}},
  \bibinfo{pages}{3815} (\bibinfo{year}{1998}).

\bibitem{Guinea98a}
\bibinfo{author}{\bibfnamefont{F.}~\bibnamefont{Guinea}},
  \bibinfo{journal}{Phys. Rev. B} \textbf{\bibinfo{volume}{58}},
  \bibinfo{pages}{9212} (\bibinfo{year}{1998}).

\bibitem{Zhang98}
\bibinfo{author}{\bibfnamefont{J.}~\bibnamefont{Zhang}} \bibnamefont{and}
  \bibinfo{author}{\bibfnamefont{R.~M.} \bibnamefont{White}},
  \bibinfo{journal}{J. Appl. Phys.} \textbf{\bibinfo{volume}{83}},
  \bibinfo{pages}{6512} (\bibinfo{year}{1998}).

\bibitem{Marx95a}
\bibinfo{author}{\bibfnamefont{A.}~\bibnamefont{Marx}},
  \bibinfo{author}{\bibfnamefont{U.}~\bibnamefont{Fath}},
  \bibinfo{author}{\bibfnamefont{W.}~\bibnamefont{Ludwig}},
  \bibinfo{author}{\bibfnamefont{R.}~\bibnamefont{Gross}}, \bibnamefont{and}
  \bibinfo{author}{\bibfnamefont{T.}~\bibnamefont{Amrein}},
  \bibinfo{journal}{Phys. Rev. B} \textbf{\bibinfo{volume}{51}},
  \bibinfo{pages}{6735} (\bibinfo{year}{1995}).

\bibitem{Marx99}
\bibinfo{author}{\bibfnamefont{A.}~\bibnamefont{Marx}},
  \bibinfo{author}{\bibfnamefont{L.}~\bibnamefont{Alff}}, \bibnamefont{and}
  \bibinfo{author}{\bibfnamefont{R.}~\bibnamefont{Gross}},
  \bibinfo{journal}{Appl. Supercond.} \textbf{\bibinfo{volume}{6}},
  \bibinfo{pages}{621} (\bibinfo{year}{1999}).

\bibitem{Kemen99}
\bibinfo{author}{\bibfnamefont{T.}~\bibnamefont{Kemen}},
  \bibinfo{author}{\bibfnamefont{A.}~\bibnamefont{Marx}},
  \bibinfo{author}{\bibfnamefont{L.}~\bibnamefont{Alff}},
  \bibinfo{author}{\bibfnamefont{D.}~\bibnamefont{K{\"o}lle}},
  \bibnamefont{and} \bibinfo{author}{\bibfnamefont{R.}~\bibnamefont{Gross}},
  \bibinfo{journal}{IEEE Trans. Appl. Supercond.} \textbf{\bibinfo{volume}{9}},
  \bibinfo{pages}{3982} (\bibinfo{year}{1999}).

\bibitem{Alers96}
\bibinfo{author}{\bibfnamefont{G.~B.} \bibnamefont{Alers}},
  \bibinfo{author}{\bibfnamefont{A.~P.} \bibnamefont{Ramirez}},
  \bibnamefont{and} \bibinfo{author}{\bibfnamefont{S.}~\bibnamefont{Jin}},
  \bibinfo{journal}{Appl. Phys. Lett.} \textbf{\bibinfo{volume}{68}},
  \bibinfo{pages}{3644} (\bibinfo{year}{1996}).

\bibitem{Lisauskas99}
\bibinfo{author}{\bibfnamefont{A.}~\bibnamefont{Lisauskas}},
  \bibinfo{author}{\bibfnamefont{S.~I.} \bibnamefont{Khartsev}},
  \bibnamefont{and} \bibinfo{author}{\bibfnamefont{A.~M.}
  \bibnamefont{Grishin}}, \bibinfo{journal}{J. Low Temp. Phys.}
  \textbf{\bibinfo{volume}{117}}, \bibinfo{pages}{1647} (\bibinfo{year}{1999}).

\bibitem{Hardner97}
\bibinfo{author}{\bibfnamefont{H.~T.} \bibnamefont{Hardner}},
  \bibinfo{author}{\bibfnamefont{M.~B.} \bibnamefont{Weissman}},
  \bibinfo{author}{\bibfnamefont{M.}~\bibnamefont{Jaime}},
  \bibinfo{author}{\bibfnamefont{R.~E.} \bibnamefont{Treece}},
  \bibinfo{author}{\bibfnamefont{P.~C.} \bibnamefont{Dorsey}},
  \bibinfo{author}{\bibfnamefont{J.~S.} \bibnamefont{Horwitz}},
  \bibnamefont{and} \bibinfo{author}{\bibfnamefont{D.~B.}
  \bibnamefont{Chrisey}}, \bibinfo{journal}{J. Appl. Phys.}
  \textbf{\bibinfo{volume}{81}}, \bibinfo{pages}{272} (\bibinfo{year}{1997}).

\bibitem{Rajeswari96}
\bibinfo{author}{\bibfnamefont{M.}~\bibnamefont{Rajeswari}},
  \bibinfo{author}{\bibfnamefont{A.}~\bibnamefont{Goyal}},
  \bibinfo{author}{\bibfnamefont{A.~K.} \bibnamefont{Raychaudhuri}},
  \bibinfo{author}{\bibfnamefont{M.~C.} \bibnamefont{Robson}},
  \bibinfo{author}{\bibfnamefont{G.~C.} \bibnamefont{Xiong}},
  \bibinfo{author}{\bibfnamefont{C.}~\bibnamefont{Kwon}},
  \bibinfo{author}{\bibfnamefont{R.}~\bibnamefont{Ramesh}},
  \bibinfo{author}{\bibfnamefont{R.~L.} \bibnamefont{Greene}},
  \bibinfo{author}{\bibfnamefont{T.}~\bibnamefont{Venkatesan}},
  \bibnamefont{and} \bibinfo{author}{\bibfnamefont{S.}~\bibnamefont{Lakeou}},
  \bibinfo{journal}{Appl. Phys. Lett.} \textbf{\bibinfo{volume}{69}},
  \bibinfo{pages}{851} (\bibinfo{year}{1996}).

\bibitem{Rajeswari98}
\bibinfo{author}{\bibfnamefont{M.}~\bibnamefont{Rajeswari}},
  \bibinfo{author}{\bibfnamefont{R.}~\bibnamefont{Shreekala}},
  \bibinfo{author}{\bibfnamefont{A.}~\bibnamefont{Goyal}},
  \bibinfo{author}{\bibfnamefont{S.~E.} \bibnamefont{Lofland}},
  \bibinfo{author}{\bibfnamefont{S.~M.} \bibnamefont{Bhagat}},
  \bibinfo{author}{\bibfnamefont{K.}~\bibnamefont{Ghosh}},
  \bibinfo{author}{\bibfnamefont{R.~P.} \bibnamefont{Sharma}},
  \bibinfo{author}{\bibfnamefont{R.~L.} \bibnamefont{Greene}},
  \bibinfo{author}{\bibfnamefont{R.}~\bibnamefont{Ramesh}},
  \bibinfo{author}{\bibfnamefont{T.}~\bibnamefont{Venkatesan}},
  \bibnamefont{and}
  \bibinfo{author}{\bibfnamefont{T.}~\bibnamefont{Boettcher}},
  \bibinfo{journal}{Appl. Phys. Lett.} \textbf{\bibinfo{volume}{73}},
  \bibinfo{pages}{2672} (\bibinfo{year}{1998}).

\bibitem{Podzorov00}
\bibinfo{author}{\bibfnamefont{V.}~\bibnamefont{Podzorov}},
  \bibinfo{author}{\bibfnamefont{M.}~\bibnamefont{Uehara}},
  \bibinfo{author}{\bibfnamefont{M.~E.} \bibnamefont{Gershenson}},
  \bibinfo{author}{\bibfnamefont{T.~Y.} \bibnamefont{Koo}}, \bibnamefont{and}
  \bibinfo{author}{\bibfnamefont{S.-W.} \bibnamefont{Cheong}},
  \bibinfo{journal}{Phys. Rev. B} \textbf{\bibinfo{volume}{61}},
  \bibinfo{pages}{3784} (\bibinfo{year}{2000}).

\bibitem{Anane00}
\bibinfo{author}{\bibfnamefont{A.}~\bibnamefont{Anane}},
  \bibinfo{author}{\bibfnamefont{B.}~\bibnamefont{Raquet}},
  \bibinfo{author}{\bibfnamefont{S.}~\bibnamefont{{von Moln\'ar}}},
  \bibinfo{author}{\bibfnamefont{L.}~\bibnamefont{{Pinsard-Godart}}},
  \bibnamefont{and}
  \bibinfo{author}{\bibfnamefont{A.}~\bibnamefont{Revcolevschi}},
  \bibinfo{journal}{J. Appl. Phys.} \textbf{\bibinfo{volume}{87}},
  \bibinfo{pages}{5025} (\bibinfo{year}{2000}).

\bibitem{Reutler00}
\bibinfo{author}{\bibfnamefont{P.}~\bibnamefont{Reutler}},
  \bibinfo{author}{\bibfnamefont{A.}~\bibnamefont{Bensaid}},
  \bibinfo{author}{\bibfnamefont{F.}~\bibnamefont{Herbstritt}},
  \bibinfo{author}{\bibfnamefont{C.}~\bibnamefont{H{\"o}fener}},
  \bibinfo{author}{\bibfnamefont{A.}~\bibnamefont{Marx}}, \bibnamefont{and}
  \bibinfo{author}{\bibfnamefont{R.}~\bibnamefont{Gross}},
  \bibinfo{journal}{Phys. Rev. B} \textbf{\bibinfo{volume}{62}},
  \bibinfo{pages}{11619} (\bibinfo{year}{2000}).

\bibitem{Palanisami01}
\bibinfo{author}{\bibfnamefont{A.}~\bibnamefont{Palanisami}},
  \bibinfo{author}{\bibfnamefont{R.~D.} \bibnamefont{Merithew}},
  \bibinfo{author}{\bibfnamefont{M.~B.} \bibnamefont{Weissman}},
  \bibnamefont{and} \bibinfo{author}{\bibfnamefont{J.~N.}
  \bibnamefont{Eckstein}}, \bibinfo{journal}{Phys. Rev. B}
  \textbf{\bibinfo{volume}{64}}, \bibinfo{pages}{132406}
  (\bibinfo{year}{2001}).

\bibitem{Raquet00}
\bibinfo{author}{\bibfnamefont{B.}~\bibnamefont{Raquet}},
  \bibinfo{author}{\bibfnamefont{A.}~\bibnamefont{Anane}},
  \bibinfo{author}{\bibfnamefont{S.}~\bibnamefont{Wirth}},
  \bibinfo{author}{\bibfnamefont{P.}~\bibnamefont{Xiong}}, \bibnamefont{and}
  \bibinfo{author}{\bibfnamefont{S.}~\bibnamefont{{von Moln\'ar}}},
  \bibinfo{journal}{Phys. Rev. Lett.} \textbf{\bibinfo{volume}{84}},
  \bibinfo{pages}{4485} (\bibinfo{year}{2000}).

\bibitem{Merithew00}
\bibinfo{author}{\bibfnamefont{R.~D.} \bibnamefont{Merithew}},
  \bibinfo{author}{\bibfnamefont{M.~B.} \bibnamefont{Weissman}},
  \bibinfo{author}{\bibfnamefont{F.~M.} \bibnamefont{Hess}},
  \bibinfo{author}{\bibfnamefont{P.}~\bibnamefont{Spradling}},
  \bibinfo{author}{\bibfnamefont{E.~R.} \bibnamefont{Nowak}},
  \bibinfo{author}{\bibfnamefont{J.}~\bibnamefont{{O\'{}Donnell}}},
  \bibinfo{author}{\bibfnamefont{J.~N.} \bibnamefont{Eckstein}},
  \bibinfo{author}{\bibfnamefont{Y.}~\bibnamefont{Tokura}}, \bibnamefont{and}
  \bibinfo{author}{\bibfnamefont{Y.}~\bibnamefont{Tomioka}},
  \bibinfo{journal}{Phys. Rev. Lett.} \textbf{\bibinfo{volume}{84}},
  \bibinfo{pages}{3442} (\bibinfo{year}{2000}).

\bibitem{Hess01}
\bibinfo{author}{\bibfnamefont{F.~M.} \bibnamefont{Hess}},
  \bibinfo{author}{\bibfnamefont{R.~D.} \bibnamefont{Merithew}},
  \bibinfo{author}{\bibfnamefont{M.~B.} \bibnamefont{Weissman}},
  \bibinfo{author}{\bibfnamefont{Y.}~\bibnamefont{Tokura}}, \bibnamefont{and}
  \bibinfo{author}{\bibfnamefont{Y.}~\bibnamefont{Tomioka}},
  \bibinfo{journal}{Phys. Rev. B} \textbf{\bibinfo{volume}{63}},
  \bibinfo{pages}{180408} (\bibinfo{year}{2001}).

\bibitem{Mathieu01}
\bibinfo{author}{\bibfnamefont{R.}~\bibnamefont{Mathieu}},
  \bibinfo{author}{\bibfnamefont{P.}~\bibnamefont{Svedlindh}},
  \bibinfo{author}{\bibfnamefont{R.}~\bibnamefont{Gunnarson}},
  \bibnamefont{and} \bibinfo{author}{\bibfnamefont{Z.~G.}
  \bibnamefont{Ivanov}}, \bibinfo{journal}{Phys. Rev. B}
  \textbf{\bibinfo{volume}{63}}, \bibinfo{pages}{132407}
  (\bibinfo{year}{2001}).

\bibitem{Glazman88}
\bibinfo{author}{\bibfnamefont{L.~I.} \bibnamefont{Glazman}} \bibnamefont{and}
  \bibinfo{author}{\bibfnamefont{K.~A.} \bibnamefont{Matveev}},
  \bibinfo{journal}{Sov. Phys. JETP} \textbf{\bibinfo{volume}{67}},
  \bibinfo{pages}{1276} (\bibinfo{year}{1988}).

\bibitem{Gross94}
\bibinfo{author}{\bibfnamefont{R.}~\bibnamefont{Gross}}, in
  \emph{\bibinfo{booktitle}{Interfaces in Superconducting Systems}}, edited by
  \bibinfo{editor}{\bibfnamefont{S.~L.} \bibnamefont{Shinde}} \bibnamefont{and}
  \bibinfo{editor}{\bibfnamefont{D.}~\bibnamefont{Rudman}}
  (\bibinfo{publisher}{Springer}, \bibinfo{address}{New York},
  \bibinfo{year}{1994}), pp. \bibinfo{pages}{176--209}.

\bibitem{Gross97}
\bibinfo{author}{\bibfnamefont{R.}~\bibnamefont{Gross}},
  \bibinfo{author}{\bibfnamefont{L.}~\bibnamefont{Alff}},
  \bibinfo{author}{\bibfnamefont{A.}~\bibnamefont{Beck}},
  \bibinfo{author}{\bibfnamefont{O.~M.} \bibnamefont{Froehlich}},
  \bibinfo{author}{\bibfnamefont{D.}~\bibnamefont{Koelle}}, \bibnamefont{and}
  \bibinfo{author}{\bibfnamefont{A.}~\bibnamefont{Marx}},
  \bibinfo{journal}{IEEE Trans. Appl. Supercond.} \textbf{\bibinfo{volume}{7}},
  \bibinfo{pages}{2929} (\bibinfo{year}{1997}).

\bibitem{Chisholm91a}
\bibinfo{author}{\bibfnamefont{M.~F.} \bibnamefont{Chisholm}} \bibnamefont{and}
  \bibinfo{author}{\bibfnamefont{S.~J.} \bibnamefont{Pennycook}},
  \bibinfo{journal}{Nature} \textbf{\bibinfo{volume}{351}}, \bibinfo{pages}{47}
  (\bibinfo{year}{1991}).

\bibitem{Kabius94a}
\bibinfo{author}{\bibfnamefont{B.}~\bibnamefont{Kabius}},
  \bibinfo{author}{\bibfnamefont{J.~W.} \bibnamefont{Seo}},
  \bibinfo{author}{\bibfnamefont{T.}~\bibnamefont{Amrein}},
  \bibinfo{author}{\bibfnamefont{U.}~\bibnamefont{Dahne}},
  \bibinfo{author}{\bibfnamefont{A.}~\bibnamefont{Scholen}},
  \bibinfo{author}{\bibfnamefont{M.}~\bibnamefont{Siegel}},
  \bibinfo{author}{\bibfnamefont{K.}~\bibnamefont{Urban}}, \bibnamefont{and}
  \bibinfo{author}{\bibfnamefont{L.}~\bibnamefont{Schultz}},
  \bibinfo{journal}{Physica C} \textbf{\bibinfo{volume}{231}},
  \bibinfo{pages}{123} (\bibinfo{year}{1994}).

\bibitem{Seo95a}
\bibinfo{author}{\bibfnamefont{J.~W.} \bibnamefont{Seo}},
  \bibinfo{author}{\bibfnamefont{B.}~\bibnamefont{Kabius}},
  \bibinfo{author}{\bibfnamefont{U.}~\bibnamefont{Dahne}},
  \bibinfo{author}{\bibfnamefont{A.}~\bibnamefont{Scholen}}, \bibnamefont{and}
  \bibinfo{author}{\bibfnamefont{K.}~\bibnamefont{Urban}},
  \bibinfo{journal}{Physica C} \textbf{\bibinfo{volume}{1995}},
  \bibinfo{pages}{25} (\bibinfo{year}{245}).

\bibitem{Mannhart96a}
\bibinfo{author}{\bibfnamefont{J.}~\bibnamefont{Mannhart}},
  \bibinfo{journal}{Supercond. Sci. Techn.} \textbf{\bibinfo{volume}{9}},
  \bibinfo{pages}{49} (\bibinfo{year}{1996}).

\bibitem{Mannhart98a}
\bibinfo{author}{\bibfnamefont{J.}~\bibnamefont{Mannhart}} \bibnamefont{and}
  \bibinfo{author}{\bibfnamefont{H.}~\bibnamefont{Hilgenkamp}},
  \bibinfo{journal}{Mater. Sci. and Eng. B} \textbf{\bibinfo{volume}{56}},
  \bibinfo{pages}{77} (\bibinfo{year}{1998}).

\bibitem{Marx97}
\bibinfo{author}{\bibfnamefont{A.}~\bibnamefont{Marx}} \bibnamefont{and}
  \bibinfo{author}{\bibfnamefont{R.}~\bibnamefont{Gross}},
  \bibinfo{journal}{Appl. Phys. Lett.} \textbf{\bibinfo{volume}{70}},
  \bibinfo{pages}{120} (\bibinfo{year}{1997}).

\bibitem{Julliere75a}
\bibinfo{author}{\bibfnamefont{M.}~\bibnamefont{{Julli\`ere}}},
  \bibinfo{journal}{Phys. Lett. A} \textbf{\bibinfo{volume}{54}},
  \bibinfo{pages}{225} (\bibinfo{year}{1975}).

\bibitem{Cabrera02}
\bibinfo{author}{\bibfnamefont{G.~G.} \bibnamefont{Cabrera}} \bibnamefont{and}
  \bibinfo{author}{\bibfnamefont{N.}~\bibnamefont{Garcia}},
  \bibinfo{journal}{Appl. Phys. Lett.} \textbf{\bibinfo{volume}{80}},
  \bibinfo{pages}{1782} (\bibinfo{year}{2002}).

\bibitem{Ziese98}
\bibinfo{author}{\bibfnamefont{M.}~\bibnamefont{Ziese}},
  \bibinfo{author}{\bibfnamefont{C.}~\bibnamefont{Srinitiwarawong}},
  \bibnamefont{and}
  \bibinfo{author}{\bibfnamefont{C.}~\bibnamefont{Shearwood}},
  \bibinfo{journal}{J. Phys.: Condens. Matter} \textbf{\bibinfo{volume}{10}},
  \bibinfo{pages}{L569} (\bibinfo{year}{1998}).

\bibitem{Ingvarsson99}
\bibinfo{author}{\bibfnamefont{S.}~\bibnamefont{Ingvarsson}},
  \bibinfo{author}{\bibfnamefont{G.}~\bibnamefont{Xiao}},
  \bibinfo{author}{\bibfnamefont{R.~A.} \bibnamefont{Wanner}},
  \bibinfo{author}{\bibfnamefont{P.}~\bibnamefont{Trouilloud}},
  \bibinfo{author}{\bibfnamefont{Y.}~\bibnamefont{Lu}},
  \bibinfo{author}{\bibfnamefont{W.~J.} \bibnamefont{Gallagher}},
  \bibinfo{author}{\bibfnamefont{A.}~\bibnamefont{Marley}},
  \bibinfo{author}{\bibfnamefont{K.~P.} \bibnamefont{Roche}}, \bibnamefont{and}
  \bibinfo{author}{\bibfnamefont{S.~S.} \bibnamefont{Parkin}},
  \bibinfo{journal}{Appl. Phys. Lett.} \textbf{\bibinfo{volume}{85}},
  \bibinfo{pages}{5270} (\bibinfo{year}{1999}).

\bibitem{Ingvarsson00}
\bibinfo{author}{\bibfnamefont{S.}~\bibnamefont{Ingvarsson}},
  \bibinfo{author}{\bibfnamefont{G.}~\bibnamefont{Xiao}},
  \bibinfo{author}{\bibfnamefont{S.~S.} \bibnamefont{Parkin}},
  \bibinfo{author}{\bibfnamefont{W.~J.} \bibnamefont{Gallagher}},
  \bibinfo{author}{\bibfnamefont{G.}~\bibnamefont{Grinstein}},
  \bibnamefont{and} \bibinfo{author}{\bibfnamefont{R.~H.} \bibnamefont{Koch}},
  \bibinfo{journal}{Phys. Rev. Lett.} \textbf{\bibinfo{volume}{85}},
  \bibinfo{pages}{3289} (\bibinfo{year}{2000}).

\bibitem{Furukawa97}
\bibinfo{author}{\bibfnamefont{N.}~\bibnamefont{Furukawa}},
  \bibinfo{journal}{J. Phys. Soc. Jpn.} \textbf{\bibinfo{volume}{66}},
  \bibinfo{pages}{2523} (\bibinfo{year}{1997}).

\bibitem{Jansen00}
\bibinfo{author}{\bibfnamefont{R.}~\bibnamefont{Jansen}} \bibnamefont{and}
  \bibinfo{author}{\bibfnamefont{J.~S.} \bibnamefont{Moodera}},
  \bibinfo{journal}{Phys. Rev. B}
  \textbf{\bibinfo{volume}{61}}(\bibinfo{number}{13}), \bibinfo{pages}{9047}
  (\bibinfo{year}{2000}).

\end{thebibliography}

\end{document}